\documentclass{iau}

\usepackage{amsmath}
\usepackage{graphicx}
\usepackage{multirow}

\newcommand{\mesa}{{\sc{mesa}}}

\newcommand{\ds}{$\delta$ Sct}
\newcommand{\gdor}{$\gamma$ Dor}
\newcommand{\dss}{$\delta$ Sct stars}

\newcommand{\rhom}{$\langle\rho\rangle$}
\newcommand{\msol}{$\mathrm{M}_{\odot}$}

\newcommand{\cd}{$\mbox{d}^{-1}$}

\newcommand{\Dnu}{$\Delta\nu$}
\newcommand{\Dnulow}{$\Delta\nu_\mathrm{low}$}
\newcommand{\etal}{\textit{et al.}}

\begin{document}

\lefttitle{Antonio Garc\'{\i}a Hern\'andez et al.}
\righttitle{The PL diagram for \dss}

\jnlPage{1}{7}
\jnlDoiYr{2021}
\doival{10.1017/xxxxx}

\aopheadtitle{Proceedings IAU Symposium}
\editors{R. de Grijs, P. Whitelock \& M. Catelan, eds.}

\title{The PL diagram for \dss: back in business as distance estimators}

\author{Antonio Garc\'{\i}a Hern\'andez$^1$, Javier Pascual-Granado$^2$, Mariel Lares-Martiz$^2$, Giovanni M. Mirouh$^1$, Juan Carlos Su\'arez$^1$, Sebastià Barceló Forteza$^1$, Andr\'es Moya$^3$}
\affiliation{
$^1$Departamento de Física Teórica y del Cosmos, Universidad de Granada, Campus de Fuentenueva s/n, 18071, Granada, Spain\\
$^2$Instituto de Astrofísica de Andalucía (CSIC). Glorieta de la Astronomía s/n. 18008, Granada, Spain\\
$^3$Departament d'Astronomia i Astrofísica, Universitat de València, C. Dr. Moliner 50, 46100, Burjassot, Spain
}
\begin{abstract}

In this work, we focus on the period-luminosity relation (PLR) of \dss, in which mode excitation and selection mechanisms are still poorly constrained, and whose structure and oscillations are affected by rotation. We review the PLRs in the recent literature, and add a new inference from a large sample of \ds. We highlight the difficulty in identifying the fundamental mode and show that rotation-induced surface effects can impact the measured luminosities, explaining the broadening of the PLR. We derive a tight relation between the low-order large separation and the fundamental radial mode frequency (F0) that holds for rotating stars, thus paving the way towards mode identification. We show that the PLRs we obtain for different samples are compatible with each other and with the recent literature, and with most observed \dss\ when taking rotation effects into account. We also find that the highest-amplitude peak in the frequency spectrum corresponds to the fundamental mode in most \ds, thus shedding some light on their elusive mode selection mechanism.
\end{abstract}


\maketitle

\section{Introduction}
\label{sec:intro}

Period-luminosity relations (PLRs) have been studied since the early twentieth century. The discovery of such a relation for classical Cepheids \citep{henrietta1908} allowed to estimate distances to other galaxies and changed our vision of the scale of the Universe. Inspired by this success, complementary PLRs have been searched for other pulsating stars, mainly those in the Cepheid instability strip \citep[see, e.g.,][and references therein]{Balona2010, Owens2022}. In particular, a PLR for \dss\ has been established since \cite{Fernie1964}.\par

\dss\ offer several observational advantages compared to Cepheids. One is their shorter periods, which arise from their smaller radii. Furthermore, as pointed out by \cite{Breger1979}, \dss\ constitute the second most abundant group of pulsators in the Galaxy, following the pulsating white dwarfs. Having a tight PLR, along with these favourable characteristics, would enable us to compare between observations and pulsation theory. Additionally, it would facilitate the determination of stellar luminosities and distances to stars, open and globular clusters, nearby galaxies, and the Galactic centre.\par

However, there are several difficulties in the interpretation of the pulsation spectra of \dss. They have spectral types from late A to early F, with masses in the range 1.5\,\msol~to 2.5\,\msol, and can be found either on the main sequence or at the H-shell burning phase. They show low-order radial and non-radial modes near the fundamental radial mode (whose frequency and period we denote as F0 and P0, respectively), although the latter is not always the highest amplitude peak in the spectrum \citep{Aerts2010}. This is one of the main difficulties to get a proper PLR, which requires the order of the radial mode to be properly identified. Moreover, other difficulties are related to their usually rapid rotation \citep{Royer2007} and the poor understanding of their excitation and selection mechanisms \citep{Aerts2010}.\par

\subsection{Recent advances in understanding the \ds\ pulsation spectrum}
\label{subsec:state of the art}

The pulsation spectrum of solar-like pulsators, including red giants, is easily understood thanks to the periodic patterns they exhibit because their pressure (p) modes appear in the asymptotic regime \cite[$n\gg\ell$, see, for example,][]{Aerts2010,corsaro2012}. The most important pattern is known as large separation, \Dnu, a spacing between consecutive radial orders of the same spherical degree that is a direct measurement of the stellar mean density.\par

\dss\ do not pulsate in the asymptotic regime but in the low-order regime of p modes. Nonetheless, prior to the launch of space missions dedicated to ultra-precise photometric time-series, several efforts were made to find a large separation structure in the pulsation spectrum of these stars \citep{handler1997, Breger99}. 
With the advent of space missions like Microvariability and Oscillations of STars \citep[{\em MOST};][]{Walker2003}, Convection, Rotation and planetary Transits \citep[{\em CoRoT};][]{auvergne2009}, and {\em Kepler} \citep{Koch2010}, it became clear that a similar pattern in the low-order regime could be found in the pulsation spectra of $\delta$ Scuti stars \citep{garciahernandez2009, garciahernandez2013, Zwintz2011,  Paparo2016sample, Sebastia2017, Michel2017, Bedding2020, Hasanzadeh2021}.\par

Theoretical predictions for these patterns were also made \citep{Reese2008, Ouazzani2015} and were found to be consistent with a large separation scaling with the mean density of the star \cite[\Dnulow\ $-$ \rhom,][]{suarez2014}. This scaling relation differs with respect to the solar-like one because of the different regimes where the modes appear in both pulsating classes. Empirical confirmation of this scaling law was subsequently provided by \cite{GarciaHernandez2015, GarciaHernandez2017} using eclipsing binary systems with a $\delta$ Scuti component. Interestingly, \cite{mirouh2019} employed a convolutional neural network to identify island modes in 2D pulsation models, and obtained a similar \Dnulow\ $-$ \rhom\ scaling relation for fast-rotating stars. The most recent theoretical relation was derived by \cite{Rodriguez-Martin2020} using synthetic spectra of 1D rotating models. The independence of the scaling relation with rotation makes \Dnulow\ a promising seismic index to shed light on the pulsation behaviour of \dss.\par

\subsection{Previous PLRs for \dss}
\label{subsec:previous}

The PLR of \dss\ has been extensively studied in previous works. \citet{McNamara2011} investigated the PLR using a sample of 26 high-amplitude $\delta$ Scuti (HADS) stars. \cite{Cohen2012b} examined 77 SX Phoenicis (SX Phe) stars, which are generally considered the Population II analogues of \dss, to explore the PLR in that class. \cite{Ziaali2019a} analysed a larger dataset consisting of 1352 {\em Kepler} \dss\ and utilised additional information from \cite{Rodriguez2001}.\par

\cite{Jayasinghe2020} conducted a study on approximately 4000 \dss\ sourced from the {\em ASAS-SN} catalogue \citep{Jayasinghe2018}. \cite{Poro2021a} focused on 27 \ds\ from the Kourovka Planet Search (KPS) project. \cite{Barac2022} investigated 434 \dss, comprising 301 from \cite{Rodriguez2001} and 133 from \citet{Chang2013}, using data obtained from the Transiting Exoplanet Survey Satellite \citep[{\em TESS};][]{Ricker2014a}.\par

\cite{DeRidder2022Gaia} studied a significant sample size of 6511 {\em Gaia} \dss\ and found evidence of a broken PLR within this population. \cite{Martinez-Vazquez2022} examined 3799 extragalactic \ds\ from various sources, including the Super MAssive Compact Halo Object \citep[{\em SuperMACHO};][]{Garg2010} and the Optical Gravitational Lensing Experiment-III \citep[{\em OGLE-III};][]{Poleski2010} projects. They also observed a broken PLR for these extragalactic objects.\par

Furthermore, \cite{Deka2022} studied 3202 \dss\ in the Galactic bulge from the {\em OGLE-IV} catalogue \citep{Soszynski2021} and the Large Magellanic Cloud (LMC) from {\em OGLE-III}. Their analysis revealed a broken PLR for the bulge \dss, but not for those in the LMC.\par

Despite the numerous studies conducted, many unanswered questions persist, and a definitive PLR for \dss\ is yet to be established. This study aims to introduce a novel PLR and utilises rotating equilibrium and pulsation models to demonstrate that rotation (through seismic and gravity darkening effects) contributes to the solution of the unresolved issues affecting the $\delta$~Scuti PLR.

\section{The samples}
\label{sec:samples}

\begin{figure}
    \centering
    \includegraphics[scale=.7]{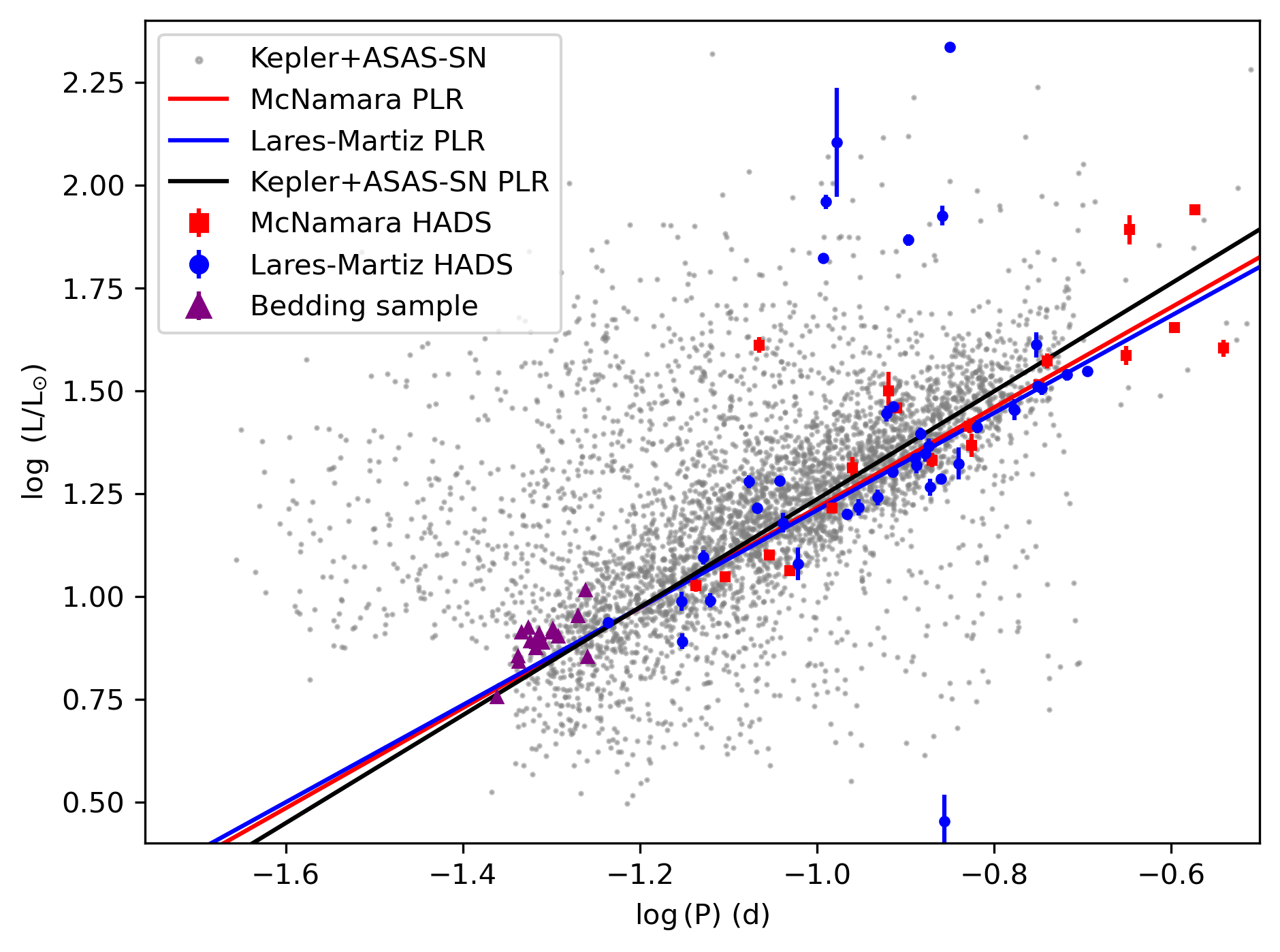}
    \caption{PLRs of the samples used in this study after removing outliers: $\delta$ Scuti, $\gamma$ Dor and hybrid stars from {\em Kepler} combined with {\em ASAS-SN} catalogue; McNamara's HADS; Lares-Martiz's HADS; and stars from \cite{Bedding2020}. Each sample is represented with a different colour and symbol. Three different PLRs are shown corresponding to the first three samples.}
    \label{fig:plr}
\end{figure}

In our study, we utilised various samples of \dss\ to investigate the PLR. We employed approximately 3700 \dss\ from the {\em ASAS-SN} catalogue, as described by \cite{Jayasinghe2020}. To ensure these objects were \ds, we selected only those stars that fell within the effective temperature range of $T_{\rm eff} = [6700, 10000] \, {\rm K}$ and surface gravity range of $\log g = [3.5, 4.2] \, {\rm dex}$, using the parameters derived from their own analysis. Additionally, we incorporated around 3800 \ds, \gdor, and hybrid stars from Sódor et al. (in prep.), obtained from the {\em Kepler} mission. We ended up discarding the pure \gdor\ stars because they are not relevant for this work.\par

We also considered a sample of 17 HADS stars with {\em Gaia} luminosities from data release 3 \citep[DR3,][]{Fouesneau2022} based on the work of \cite{McNamara2011}. We added a new sample of HADS composed of 11 HADS stars with non-distorted lightcurves from \cite{Lares-Martiz2022} and 36 HADS stars introduced in this work for the first time, collectively forming what we refer to as Lares-Martiz's HADS sample. These 47 lightcurves were acquired from the {\em TESS} satellite. The Best Parent Method \citep{Lares-Martiz2020} was employed to determine their F0 frequency, which is usually that of the dominant peak for HADS. By fitting not only the main peaks but also their possible combinations, this method ensures a robust and more precise determination of F0. The list and F0 values of the stars in this sample are shown in Table\,\ref{tab:table_mlmHADS}.\par

\begin{table}
\begin{center}
\caption{List of HADS comprising the Lares-Martiz sample. The 11 HADS taken from \cite{Lares-Martiz2022} are indicated by a star symbol preceding their {\em TESS} Input Catalog (TIC) ID, while the remaining 36 are presented here for the first time. F0 was extracted from {\em TESS} data using the Best Parent Method \citep{Lares-Martiz2020}.}
\label{tab:table_mlmHADS}
  \begin{tabular}{rrcrrcrr}
  \midrule

TIC ID & F0 [d$^{-1}$] &
&
TIC ID & F0 [d$^{-1}$] &
&
TIC ID & F0 [d$^{-1}$] \\

\midrule
  17153995 & 8.54481 & %
& 181087723 & 7.72287 &
& 334436767 & 7.47803 \\
  35920894 & 5.64856 & %
& $\star$ 183532876 & 5.98212 &
& 339675323 & 7.44527 \\
  43173526 & 11.93649 & %
& 200624064 & 5.2087  &
& 342500794 & 6.5115  \\
  46937596 & 7.88982 & %
& 210548440 & 5.42033 &
& $\star$ 355547586 & 10.51244 \\
  78850814 & 9.11811 & %
& $\star$ 224285325 & 18.19356 &
& $\star$ 355687188 & 8.19196 \\
$\star$  51991595 & 10.91561 & %
& $\star$ 231632224 & 6.91544 &
& $\star$ 358502706 & 11.67333\\
  80781425 & 6.5968 & %
& 241787384 & 11.0103 &
& 362384415 & 7.21626 \\
 118080601 & 7.54064 & %
& 241843363 & 5.98816 &
& 374753270 & 7.24599 \\
$\star$ 121731704 & 14.22167 & %
& 242302902 & 7.63725 &
& 383604347 & 9.49098 \\
$\star$ 139845816 & 6.80218 & %
& 262652067 & 5.63467 &
& 393420032 & 6.01012 \\
$\star$ 144309524 & 8.97501 & %
& 293110952 & 14.20631 &
& 396424970 & 7.1762  \\
 145372195 & 7.74984 & %
& 298112357 & 9.84591 &
& 447363101 & 5.22441 \\
$\star$ 166979292 & 17.217331 & %
& 299948201 & 7.07711 &
& 448892817 & 13.43279 \\
 168384036 & 9.25355 & %
& 304196197 & 8.82052 &
& 454665792 & 8.21272 \\
 178463456 & 4.94822 & %
& 308396022 & 13.20399&
& 464401254 & 8.33771 \\
 178616716 & 9.76366 & %
& 321528595 & 5.57497 &
& & \\

\midrule
\end{tabular}
\end{center}
\end{table}

Finally, we have used 60 young \dss\ identified by \cite{Bedding2020} in our analysis. This sample has the advantage of showing clear periodicities, so \Dnulow\ was identified.\par

To ensure consistency and obtain luminosity information, we cross-matched all the aforementioned catalogues with {\em Gaia} DR3 \citep{Fouesneau2022}. The resulting PL diagram is shown in Figure\,\ref{fig:plr}. The combined {\em ASAS-SN} and {\em Kepler} sample is shown as grey points, whereas McNamara's, Lares-Martiz's, and Bedding's samples are depicted with red squares, blue circles, and purple triangles, respectively. Uncertainties both in luminosity and period are also shown for the smaller sample (for the sake of clarity), but they are usually smaller than the symbol size. The entire \ds\ sample spans approximately $\log P \simeq [-1.6, -0.7]$ and $\log (L/L_\odot) \simeq [0.75, 2.00]$ . Within this region, a denser band was observed, primarily consisting of stars where the fundamental mode (of period P0) also exhibited the highest amplitude \citep{Jayasinghe2020}. This denser band was used to calculate the PLR, as described in Section~\ref{sec:PLR}.\par

\section{Getting the PLR}
\label{sec:PLR}

We have obtained three different PLRs with the previous samples that are depicted in Figure\,\ref{fig:plr}. We obtain a PLR using Lares-Martiz's sample of HADS. We avoid outliers in the fit (more than $2\sigma$ of deviation from the fit). Some outliers seem to be binary systems, for example, TIC\,299948201, identified as a double or multiple-star system in Simbad \citep{Wenger2022}. Other outliers suspected to be binaries are TIC\,383604347 and TIC\,396424970. Initially, these  were identified as \dss\ by \cite{Chang2013}, but recently \cite{Barac2022} identified them as non-\ds\ since binary systems without a pulsating component can also look like HADS. This can explain the high luminosity of TIC\,383604347 ($\log (L/L_\odot) > 1.75$, see Figure\,\ref{fig:plr}) and low luminosity ($\log (L/L_\odot) < 0.5$) of TIC\,396424970.\par 

We also calculated the PLR for McNamara's sample of HADS. We did not take into account AE UMa, the star with $\log (L/L_\odot) \sim 1.5$ and $\log P \sim -1.05$, because its luminosity differs from that found by other authors (for example, \citealt{Xue2022} find $\log (L/L_\odot)$ between 1.00 and 1.24).\par

Finally, to get the PLR from the combined {\em Kepler} and {\em ASAS-SN} sample, we used the denser band of stars in the PL diagram (see Section~\ref{sec:samples}). We divided the set in equal bins of $\log P$ to get 20 bins. This assured a statistically significant number of stars per bin. Then, we created a histogram in $\log (L/L_\odot)$ of every bin and calculated the Gaussian kernel density estimator to obtain its maximum value. With that maximum and the centre of the bin, we get the points to be fitted and derive the PLR.\par

The relations found in the way described above give, respectively:
  \begin{subequations}
    \begin{align}
      \log (L / L_\odot)_\textrm{Lares-Martiz} = (1.18\pm0.10)\cdot\log P + (2.39\pm0.09), \label{eq:mariel}\\
      \log (L / L_\odot)_\textrm{ASAS-SN} = (1.21\pm0.15)\cdot\log P + (2.43\pm0.13), \label{eq:asasn}\\
      \log (L / L_\odot)_\textrm{McNamara} = (1.31\pm0.03)\cdot\log P + (2.55\pm0.03). \label{eq:mcnamara}
    \end{align}
  \end{subequations}

These relations, in which the period is expressed in days, are all in agreement, within the uncertainties at $1\sigma$. The higher deviation comes from the independent term in the case of equation\,(\ref{eq:mariel}), corresponding to Lares-Martiz's sample. This is probably due to the fact that most of the stars appear in the lower part of the broad region of the {\em ASAS-SN} sample in the PL diagram. For the rest of the work, we will use equation\,(\ref{eq:asasn}), corresponding to the {\em ASAS-SN} sample, because the corresponding PLR is obtained for a larger number of stars. This statistically more robust sample allows avoiding certain problems and errors, such as wrong luminosity values. It is not surprising that this relation has the lowest error bars.\par

\begin{figure}
    \centering
    \includegraphics[scale=.7]{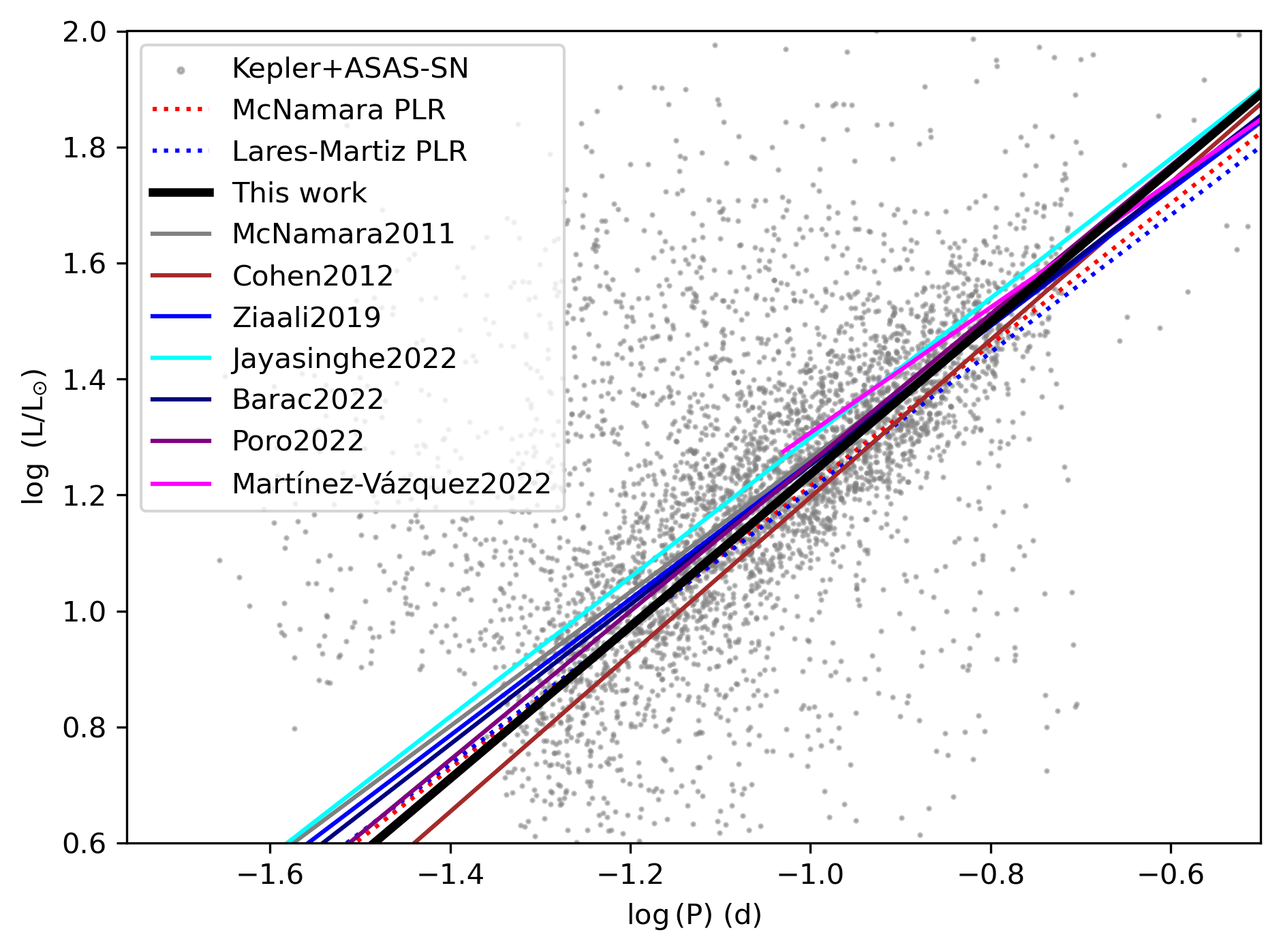}
    \caption{Comparison between PLRs found in previous studies and this work (corresponding to equation~(\ref{eq:asasn})). The PLRs of HADS stars from the McNamara's and Lares-Martiz's samples are also shown as a guide. Notice how similar the slopes of the different fits are. The short-period segment of the \citet{Martinez-Vazquez2022} relation is not depicted because of the broken relation they found, we only include the segment most coherent with the other PLRs.}
    \label{fig:previous}
\end{figure}

We also compared our findings to previous PLRs. Figure~\ref{fig:previous} shows several previous studies\footnote{Because we just used the previous PLRs for comparison reasons, we transform the visual absolute magnitudes given by the authors into bolometric ones assuming that most \ds\ are A-type stars for which the bolometric correction is 0.} together with our new relations. As can be seen, the relations obtained with the different samples in this work are essentially the same as in previous studies. Small deviations are expected due to differences in the samples used to get them.\par

\section{Gravity darkening effects due to rapid rotation}

\dss\ are rapid rotators \citep[see, e.g.,][]{Royer2007}. Due to their high rotation speeds, these stars become oblate, deviating from the shape of an ideal sphere. This deformation gives rise to distinct equilibrium conditions on their surfaces, resulting in higher temperatures and brightness at the poles, while the equator experiences lower temperatures and fainter luminosity. This phenomenon is commonly referred to as gravity darkening, and it holds significant importance when considering observed luminosities.\par

Rotation effects have commonly been incorporated into 1D codes as a correction factor for the stellar equations, as done, for instance, in the Modules for Experiments in Stellar Astrophysics \citep[\mesa;][]{Paxton2019}. In this code, after calculating the stellar structure for a specific rotation rate, a gravity darkening model derived from 2D solutions is applied \citep{EspinosaLara2011}. Inclination also affects the observed physical parameters in this case, as the relative visibility of the pole versus the equator plays a role. Under extreme conditions, the combined influence of stellar deformation and inclination can lead to a luminosity increase of up to 50\% and a temperature variation of 2.5\%. Consequently, this effect can significantly impact a PLR.\par

\begin{figure}
    \centering
    \includegraphics[scale=.7]{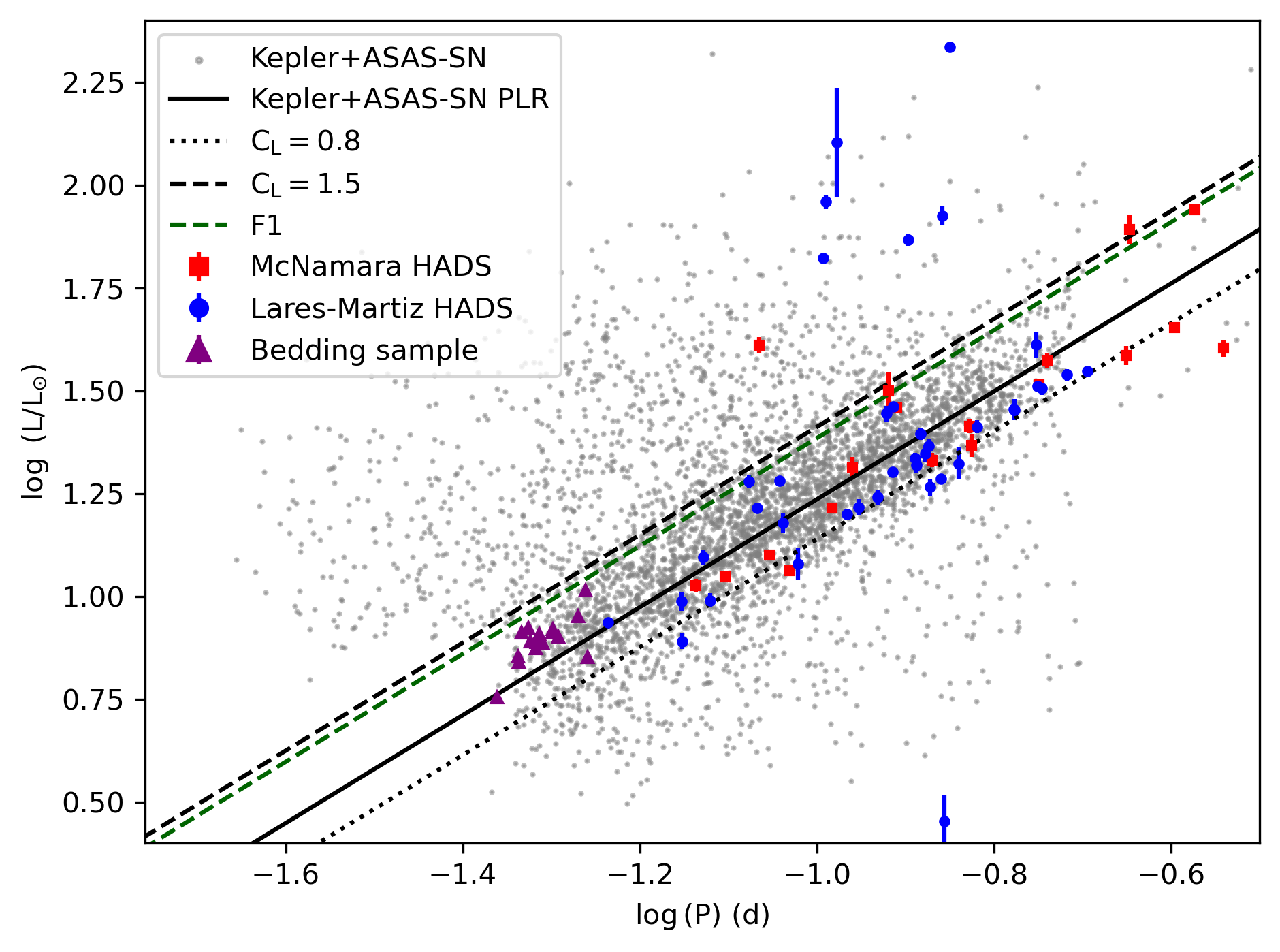}
    \caption{The PLR found in this work is shown with the upper and lower limits of projected luminosity $C_L$, the first overtone F1, and all samples of stars analysed.}
    \label{fig:gravity_darkening}
\end{figure}

To show the possible impact of not taking into account the gravity darkening, we have used the two limiting cases, i.e., the highest rotation rate possible but with the line of sight at 0$^\circ$ (pole-on) or 90$^\circ$ (equator-on). In these cases, the luminosity can be higher by up to 50\%\ or lower by up to 20\%, respectively \citep{Paxton2019}. The angle at which the projected luminosity is equal to the real luminosity is approximately 55$^\circ$, meaning that it is slightly more probable to find a star with higher projected luminosity (supposing randomly distributed inclinations). 
All this is shown in Figure\,\ref{fig:gravity_darkening}. We plot our PLR and also the limits of the projected luminosity, $C_L \equiv L_\mathrm{proj}/L$, over the diagram with the samples. Clearly, the gravity darkening can explain the broader region observed in our large sample of more than 4000 stars. Moreover, if we estimate the first overtone period (P1) based on P1/P0 $\sim$ 0.77, then we also see in the figure that there exists overlapping between the pole-on case and that overtone, depicted as a dashed green line. This translates into difficulty in getting a reliable mode identification, even for the fundamental mode. The only way to overcome this problem is to derive the unprojected stellar rotational velocity, something hard to get even with asteroseismology \cite[see, for example, ][]{Ramon-Ballesta2021}.\par

\subsection{The $F0-\Delta\nu_{low}$ relation}

\begin{figure}
    \centering
    \includegraphics[scale=.7]{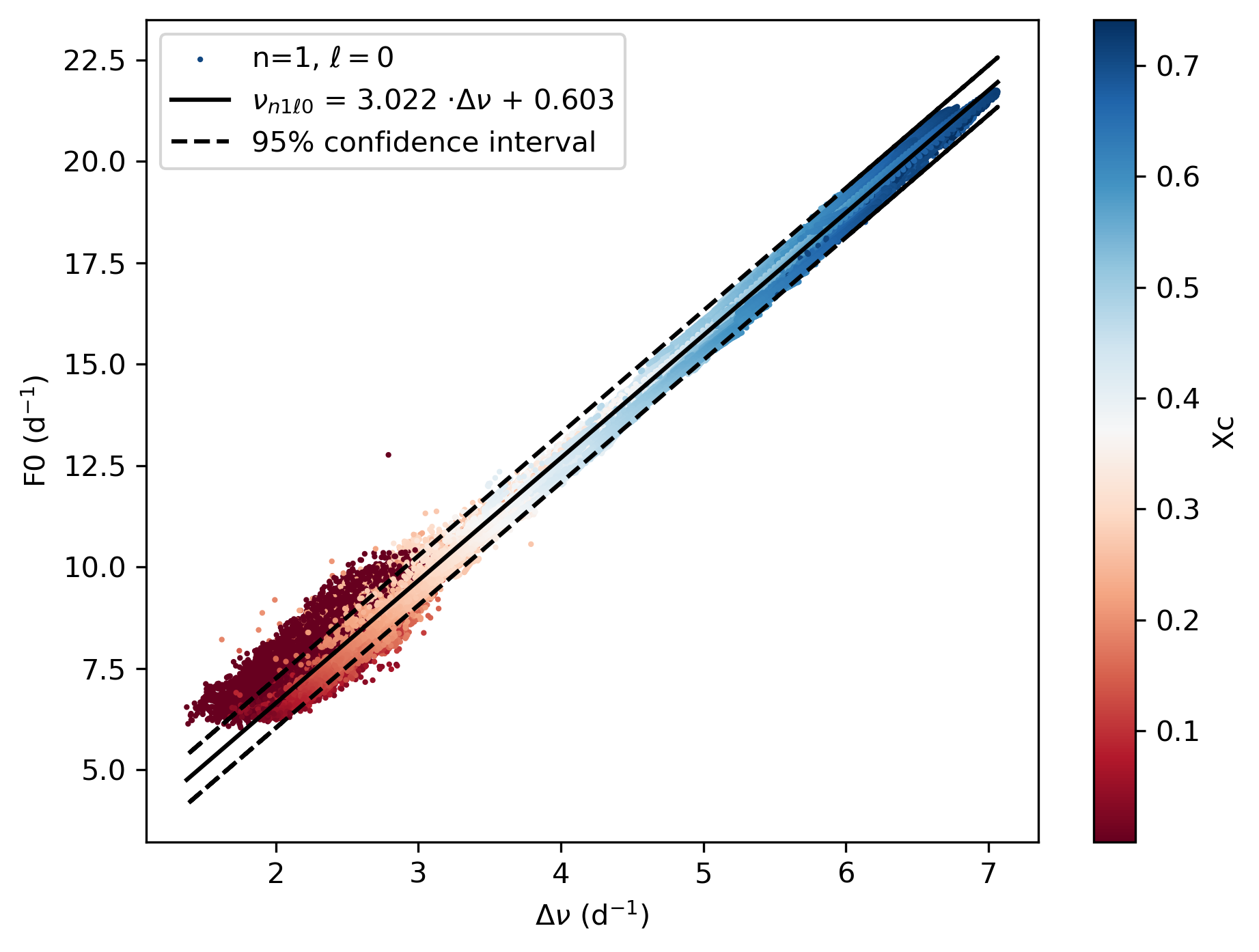}
    \caption{$\rm{F0}-\Delta\nu_{low}$ relation. The fundamental mode ($n=1$, $\ell = 0$) and the low-order large separation (\Dnulow) for rotating equilibrium and pulsation models in the observational instability strip of $\delta$ Sct stars are shown. The plot is colour-coded according to the central hydrogen abundance, Xc, illustrating the evolutionary stage of each model. A fit to this relation is shown as a continuous line, with the associated 95\% confidence interval shown as dashed lines.}
    \label{fig:F0vsDnu}
\end{figure}

Apart from dealing with gravity darkening, positioning a particular star in the PLR means a correct identification of the fundamental radial mode. Because both the low-order large separation and P0 are directly related to the stellar mean density \citep{GarciaHernandez2015,GarciaHernandez2017}, a direct relation between both quantities is also likely. To check that, we have used the rotating equilibrium and pulsation models by \cite{Rodriguez-Martin2020} to look for such a relation. Figure\,\ref{fig:F0vsDnu} shows the results. In this figure, we have plotted the fundamental mode frequency (F0) as a function of the low-order large separation (\Dnulow) for the models in the observational instability strip of \dss, as found by \cite{Murphy2019}. A tight linear fit is clearly seen, with some scatter in the most evolved stages that correspond to the lowest values of F0 and \Dnulow. The deviation from the relation is somewhat expected in the models since the stars at those stages are quite evolved, probably indicating that they are not even \ds\ pulsators. A closer inspection of these models with non-adiabatic calculations of the oscillation frequencies would be necessary to confirm their mode stability. \par
The result of the fit in \cd\ is:
\begin{equation}
      \rm{F0} = 3.022\cdot\Delta\nu_{low} + 0.603, \label{eq:DnuF0}
\end{equation}

\noindent with $r^2 = 0.995$ and a standard deviation of the residuals equal to $\sigma = 0.311$\,\cd. There have been previous experimental determinations of a $\rm{F0}-\Delta\nu_{low}$ relation \citep{Jayasinghe2020, Bedding2020}. Our fit is not very different from others, although it is much more {robust}. The difficulties in determining the correct fundamental mode and \Dnulow\ have hampered the use of this relation in the past.\par

\subsection{Towards a fundamental mode identification}

\begin{figure}
    \centering
    \includegraphics[scale=.7]{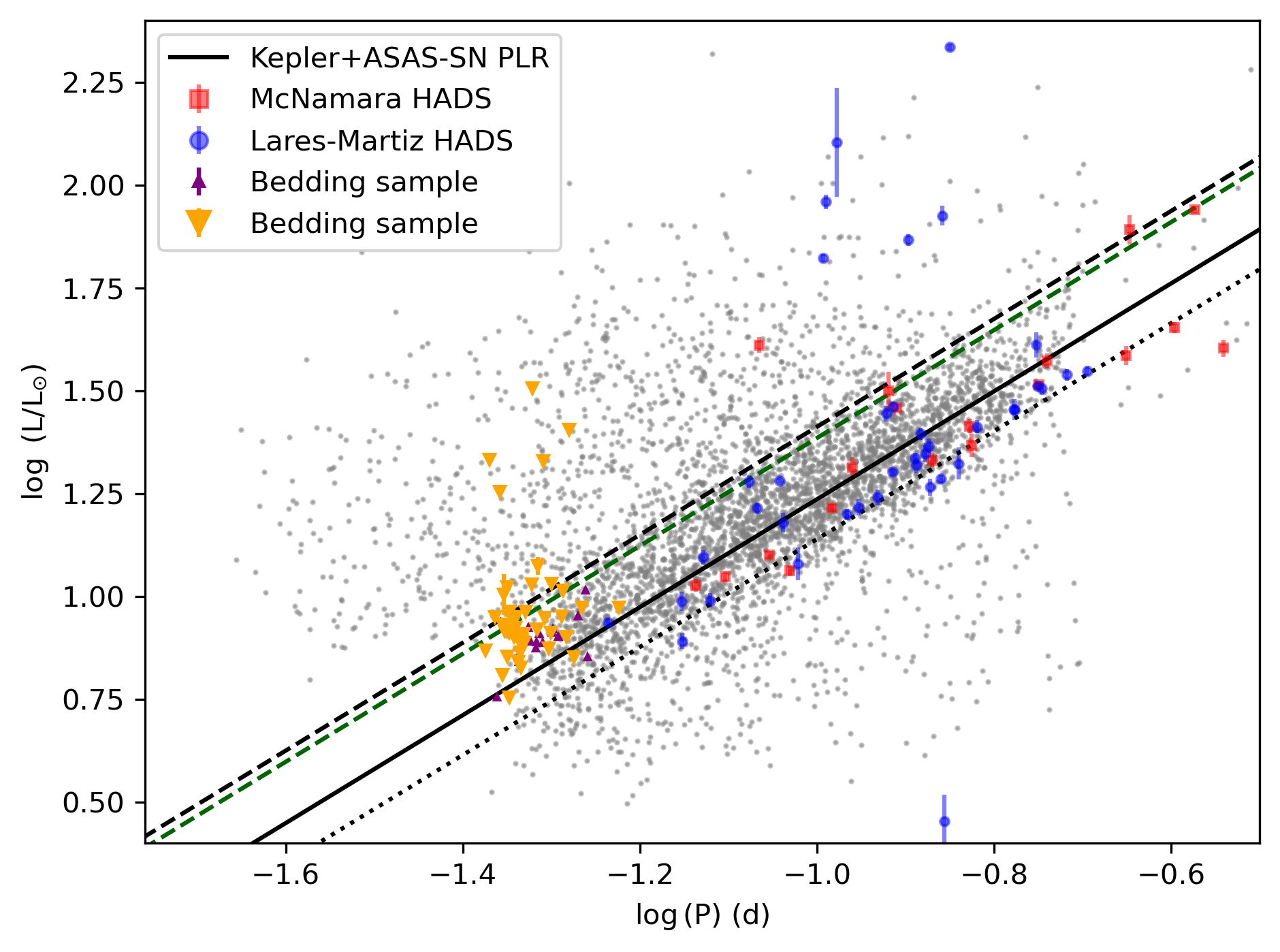}
    \caption{The PLR found in this work is shown with the upper and lower limits of the projected luminosity $C_L$ (dashed and dotted black lines, respectively), the first overtone F1 (green dashed line), and all samples of stars analysed. 
    F0 computed from the $\rm{F0}-\Delta\nu_{low}$ relation given in equation\,(\ref{eq:DnuF0}) for the \cite{Bedding2020} stars is shown with downward yellow triangles.}
    \label{fig:F0 ID}
\end{figure}

One may consider using the relations we present here to derive P0 knowing \Dnulow, but the large separation has only been determined for a handful of \ds. The larger sample to date is that of \cite{Bedding2020}. They were able to find \Dnulow\ for 60 young \dss, for which the pattern of the large separation is clearer. They also identified the fundamental mode for some of them, which are the stars depicted in Figure\,\ref{fig:plr} as purple triangles. We tried to check the validity of their mode identification and complete it for those stars without an F0 determination in \cite{Bedding2020}.\par

Using equation\,(\ref{eq:DnuF0}), we computed new F0 values based on the provided \Dnulow. Figure\,\ref{fig:F0 ID} shows a similar plot to Figure\,\ref{fig:gravity_darkening}, but with the computed F0 as yellow triangles. Most of our determined fundamental modes are in agreement with those found by \cite{Bedding2020}, with some slight differences coming from a non-perfect fit (included in the uncertainties). However, for some stars showing higher luminosities, F0 does not seem to be correctly determined: a complete description thus requires a better understanding of the selection mechanism.
While our models predict fundamental modes corresponding to luminosities above and below the PLR, observations only populate the region close to it. These results point to the impossibility of populating the PLR with new, isolated stars using just \Dnulow. However, it appears that only stars in which the fundamental mode is actually excited lie close to the PLR: if this holds true, a detailed study of the stars with a detectable fundamental mode at F0 might be of interest to shed light on the selection mechanism, which has remained evasive even with the advances permitted by recent space missions.\par

\section{Conclusions}

We have derived a new PLR for \dss\ by utilizing a sample of around 4000 such stars from {\em ASAS-SN} and {\em Kepler} data. Our relation is consistent with previous determinations. We have also brought out that gravity darkening plays a significant role in causing the broadening effect within the PLR. This broadening includes the predicted position of the first overtone, hampering the identification of both F0 and F1.\par

We have derived a theoretical $\rm{F0}-\Delta\nu_{low}$ relation that is compatible with previous observations, demonstrating its robustness even in the presence of rotation. Using this relation to guess F0 does not help to get a more constrained PLR. Only in the cases when the fundamental mode is excited (i.e., detected in the frequency spectrum) does it place the pulsator close to the PLR. This observational fact provides valuable insights into the excitation and mode selection processes.\par

There also seems to be an internal process related to mode selection that favours the high-amplitude peak being associated with the fundamental mode in the majority of \dss, as suggested by the broader region of the PL diagram that forms the PLR. This might offer valuable clues for further investigations into their excitation mechanisms and mode selection processes.

\begin{acknowledgements}
The authors acknowledge T. Bedding for allowing the use of the F0 values from his 2020 publication. AGH acknowledges funding support from ‘FEDER/junta de Andalucía-Consejería de Economía y Conocimiento’ under project E-FQM-041-UGR18 by Universidad de Granada. JPG and MLM acknowledge financial support from project PID2019-107061GB-C63 from the `Programas Estatales de Generaci\'on de Conocimiento y Fortalecimiento Cient\'ifico y Tecnol\'ogico del Sistema de I+D+i y de I+D+i Orientada a los Retos de la Sociedad' and from the grant CEX2021-001131-S funded by MCIN/AEI/10.13039/501100011033. AGH, JCS, GMM and SBF acknowledge funding support from the Spanish State Research Agency (AEI) project PID2019-107061GB-064. Funding for open access charge: Universidad de Granada.
\end{acknowledgements}

\end{document}